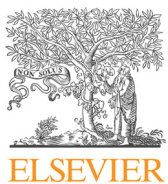
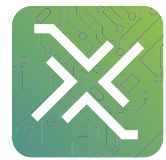

# Scan4CFU: Low-cost, open-source bacterial colony tracking over large areas and extended incubation times

Santosh Pandey [a,*], Yunsoo Park [a], Ankita Ankita [a], Gregory J. Phillips [b,*]

[a] Electrical and Computer Engineering, Iowa State University, Ames, IA, USA
[b] Veterinary Microbiology and Preventive Medicine, Iowa State University, Ames, IA, USA

ABSTRACT

A hallmark of bacterial populations cultured *in vitro* is their homogeneity of growth, where the majority of cells display identical growth rate, cell size and content. Recent insights, however, have revealed that even cells growing in exponential growth phase can be heterogeneous with respect to variables typically used to measure cell growth. Bacterial heterogeneity has important implications for how bacteria respond to environmental stresses, such as antibiotics. The phenomenon of antimicrobial persistence, for example, has been linked to a small subpopulation of cells that have entered into a state of dormancy where antibiotics are no longer effective. While methods have been developed for identifying individual non-growing cells in bacterial cultures, there has been less attention paid to how these cells may influence growth in colonies on a solid surface. In response, we have developed a low-cost, open-source platform to perform automated image capture and image analysis of bacterial colony growth on multiple nutrient agar plates simultaneously. The descriptions of the hardware and software are included, along with details about the temperature-controlled growth chamber, high-resolution scanner, and graphical interface to extract and plot the colony lag time and growth kinetics. Experiments were conducted using a wild type strain of *Escherichia coli* K12 to demonstrate the feasibility and operation of our setup. By automated tracking of bacterial growth kinetics in colonies, the system holds the potential to reveal new insights into understanding the impact of microbial heterogeneity on antibiotic resistance and persistence.

## Specifications Table

| | |
|---|---|
| Hardware name | Bacterial Growth, Imaging, and Analysis Station (Scan4CFU) |
| Subject area | • Biological Sciences (Microbiology)<br>• Environmental, Planetary and Agricultural Sciences |
| Hardware type | • Imaging tools |
| Closest commercial analog | Quebec™ Dark-Field Colony Counter, IncuCount™ Colony Counter, Corning Cell Counter, and Bel-Art™ SP Scienceware™ Plate Reader |

---

\* Corresponding authors.
  *E-mail addresses:* pandey@iastate.edu (S. Pandey), yunsoopk@iastate.edu (Y. Park), ankita@iastate.edu (A. Ankita), gregory@iastate.edu (G.J. Phillips).

*(continued)*

| Hardware name | Bacterial Growth, Imaging, and Analysis Station (Scan4CFU) |
| --- | --- |
| Open Source License | CC-BY-SA 4.0 |
| Cost of Hardware | *$306.40* |
| Source File Repository | Zenodo Data Repository Pandey, Santosh (2021), "ScanCFU: Low-cost, open-source bacterial colony tracking over large areas and extended incubation times ", Zenodo, V1, /https://doi.org/10.5281/zenodo.5636769 |

**Hardware in context**

Counting bacterial colonies that form on nutrient agar plates (i.e., plate counting) is a standard practice in microbiology in order to quantify the number of individual bacteria within a culture, and is based on reporting on the number of colony forming units (CFU)/mL of culture. However, this method is not designed to measure growth kinetics of bacteria as they form visible colonies [1]. There are multiple studies, however, that would benefit from the ability to monitor growth kinetics of bacterial colonies over prolonged time scales. For example, since bacteria growing in broth culture encounter different environmental conditions than those growing on a solid substrate, such as spatial limitations and oxygen availability, measuring the impact of these differences would be valuable. It is also now clear that even bacteria growing in exponential growth phases, under so called "balanced" growth conditions can exhibit heterogeneity with respect to growth rate, cell size and content [2–4]. Importantly, these cells have been linked to the phenomenon of antibiotic persistence, where a small subpopulation of cells reside in a state of dormancy causing them to be less susceptible to the lethal action of antibiotics [4–6]. Because persistence is associated with the emergence of antibiotic resistance and reoccurrence of bacterial infections [3], new methods are needed to understand the contribution of persister cells, and cells that have entered into an asynchronous growth phase in general, to overall bacterial growth and adaptation [7,8]. For example, the impact that cells (that are not in a state of synchronous growth) have on the formation of individual colonies is not known [7,9,10].

Automated counting of bacterial colonies has been pursued by researchers in both academics and industry to improve the experimental throughput, reproducibility, processing time, and documentation [7–12]. Commercialized products for bacterial enumeration include optical density counters (e.g., spectrophotometer), digital colony counters (e.g., Neutec Group's Sphere Flash Colony Counter), and automated image analysis systems (e.g., BD Kiestra System). Some other examples are the Quebec™ Dark-Field Colony Counter, IncuCount™ Colony Counter, Corning Cell Counter, and Bel-Art™ SP Scienceware™ Plate Reader. These commercialized products are expensive, typically priced between $1000 to $10,000 USD per system. The commercialized image processing software are usually proprietary and restrictive, and do not fully support user modification or sharing within communities.

As an alternative to commercialized products, there is a need to democratize microbiology tools for bacterial image capture and image analysis [9]. Previous researchers have recorded digital images of Petri dish using webcams, DSLR cameras, smartphones, and flatbed scanners [7]. For example, a recent study used a Canon EOS 500D 15 MP SLR Camera to record digital plate images in grayscale and RGB color at 300 and 800 dpi resolution, respectively [11]. Another recent study employed the Canon EOS 1200D reflex camera triggered by an Arduino controller to record plate images [2]. In addition, a number of open source software have been released for this application [7,9–13]. Over the years, image processing and analysis techniques have significantly refined to improve robustness, versatility, and user-friendliness. Various algorithms and plug-ins have been added to overcome the common artifacts and imperfections during image capture, background, color, contrast, edge effects, and colonies amongst agar plates. One example is the open-source ImageJ software originally developed by Wayne Rasband at the National Institutes of Health [14]. In recent years, GitHub has emerged as the public repository for open-source software for bacterial counting from images of plates, such as the DDot Counter based on the watershed algorithm [13].

While significant improvements have been made in image recognition and analysis, there is a dearth of open-source hardware for incubating bacterial colonies and tracking their growth on multiple agar plates in parallel and over several hours. In this work, we address this need by presenting an open-source station to study the growth kinetics and lag time of bacterial colonies. The image capture is conducted inside a temperature-controlled acrylic chamber by placing multiple nutrient agar plates with seeded cells on a high-resolution flatbed scanner (resolution of 2400 dpi or higher). The scanned images of every plate are automatically recorded at fixed intervals and stored in a designated folder through the course of the experiment. The image analysis is performed on the recorded images to extract various growth parameters of single colonies within each plate. A graphical user interface (GUI) is developed to navigate through the various steps of image processing. The hardware and software for image capture and image analysis are described, along with the bill of materials, and instructions for building and operating the system. Lastly, the limitations of the presented technology and future directions are discussed. The hardware and software are open-source and made available in the online Zenodo Data Repository at /https://doi.org/10.5281/zenodo.5636769

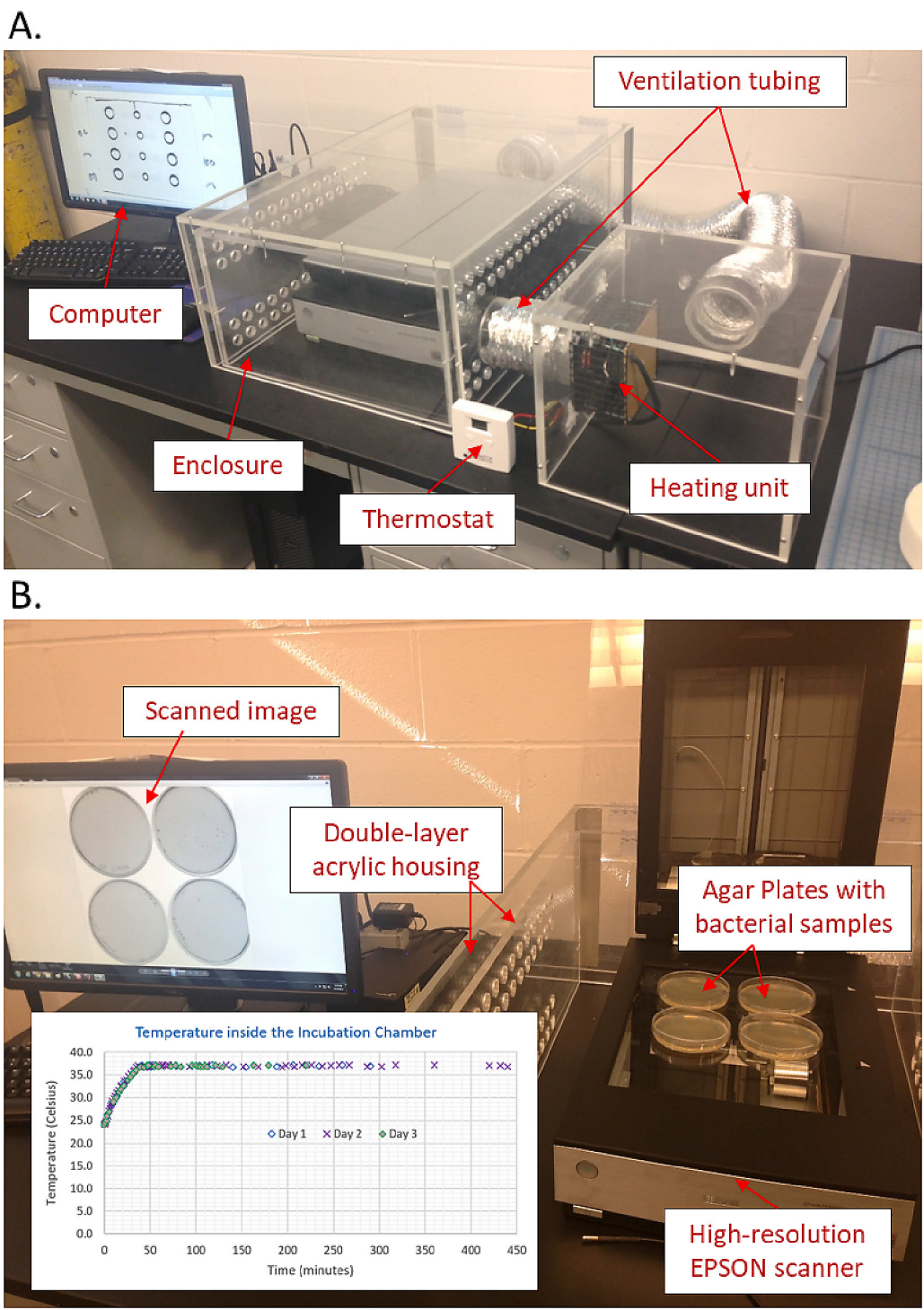

**Fig. 1.** Open-source hardware to track the growth and imaging of bacterial colonies on agar plates. (A) The enclosure for housing the scanner is shown, along with the heating unit, thermostat to monitor the internal temperature, ventilation tubing, and desktop computer. (B) Four nutrient agar plates are placed on the EPSON flatbed scanner. Scanned image of the four agar plates are displayed on the monitor. The inset shows the temperature within the enclosure as a function of time.

## Hardware description

Our bacterial growth and imaging hardware consists of two chambers (i.e., interior and exterior chambers) to enclose a high-resolution scanner, a side chamber with an attachable heater/fan unit, and a vent system to regulate the air flow through the chambers (Fig. 1). The interior chamber has 1-inch diameter holes to avoid direct air flow over the scanner. The chambers were custom-built using acrylic material, but other durable materials can also be used. The heater/fan unit was low cost and purchased from an outside vendor. Other vendors for a portable heater and fan are available at local hardware stores. The main imaging module is the Epson Perfection V750 Pro Scanner having a 6400 dpi scan resolution and a maximum scan area of 21.6 cm × 29.7 cm. A cheaper option is the Epson V19 Flatbed Scanner that has a resolution of 4800 dpi which is sufficient for the imaging experiments. The system described here builds upon our previous work on single cell imaging studies that employed an acrylic incubation chamber to house a student upright microscope [8].

For plating and growing the bacteria, nutrient agar plates (3-inch diameter) have been used (Fig. 1B). One imaging experiment can image up to six such nutrient agar plates. Other sizes of agar plates can be used, provided that all the plates fit within the maximum scanning area (21.6 cm × 29.7 cm). For plating cells on the agar surface, our goal was to have less than 250 colonies in the entire plate. This is because an appropriate colony count is between 25 and 250 *colony forming units* (CFU) per plate, according to the U.S. Food and Drug Administration (FDA) [15]. If the colonies are over 250 CFUs, the growth may be inhibited for some bacteria [15]. On the other hand, plates with less than 25 CFUs are not considered statistically representative of the test sample [15]. Each CFU on the plate is assumed to have emerged from a single cell or a small group of cells. We tried to reduce occurrence of colony clusters during initial plating of cells by appropriately shaking the cell populations and choosing a suitable dilution before seeding.

## Software description

Our custom software is designed to convert a group of scanned plate images into a video file for easy visualization, to remotely control the operation of the EPSON scanner, to effectively subtract background from the scanned plate images, to isolate and detect individual bacterial colonies, and to plot the increasing area of different colonies as a function of time. The software is written as a collection of Matlab Files with instructions in the readme file. Upon downloading and running the files according to the instructions, a new window pops up as shown in Fig. 2. A number of colonies are automatically selected with unique color codes, and some of them can be de-selected if they are touching or merging with one another. An X-Y cursor is used to select the background area. Thereafter, the background is subtracted from the images, and all the individual colonies are found and tracked by their respective area and centroid position. A plot of the tracked bacterial colony area is then generated as a function of the recording time.

## Design files

### Design files summary

The Design files correspond to the CAD files to build the acrylic chamber housings around the scanner and heater/fan unit, .exe files for automated image capture from the scanner (zip file folder: Automated Image Capture). and MatLab files (zip file

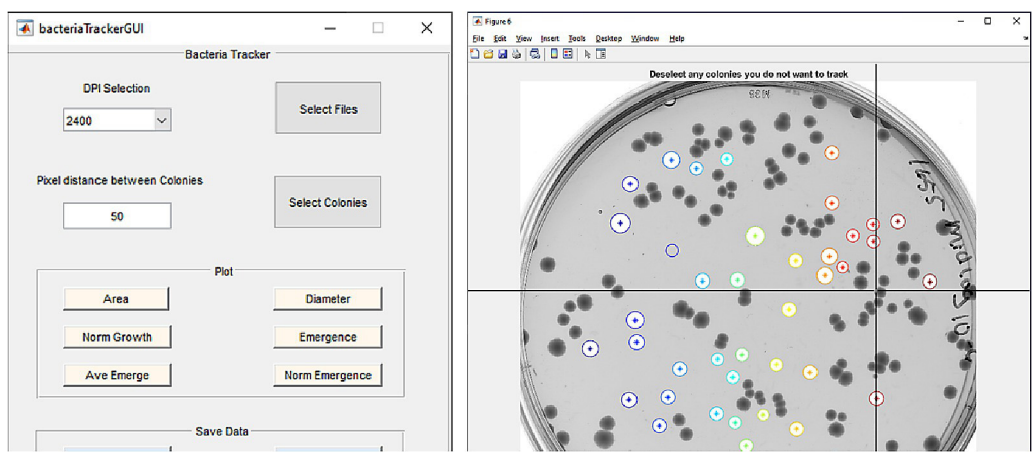

**Fig. 2.** Open-source software to track the growth and imaging of bacterial colonies on agar plates. The graphical interface is shown on the left that has clickable buttons to select image files and de-select colonies that are merged or touching one another. An X-Y cursor is used to select the background area for subsequent background subtraction and identification of bacterial colonies. Thereafter, a number of parameters can be plotted, such as colony area, diameter, and emergence. The data is exported to Microsoft Excel for further analysis. A representative image of the nutrient agar plate is shown on the right with × and y slide bars.

folder name: MATLAB_Files_for_Bacteria_Tracking) to automatically detect bacterial colonies from a sequence of images and track the colony area as a function of time. The source files mentioned below are posted in the Zenodo Data Repository, V1, / https://doi.org/10.5281/zenodo.5636769

| Design file name | File type | Open source license | Location of the file |
|---|---|---|---|
| Exterior Chamber | CAD file (.dwg) | CC-BY-SA 4.0 | available with the article |
| Interior Chamber | CAD file (.dwg) | CC-BY-SA 4.0 | available with the article |
| Heater Chamber | CAD file (.dwg) | CC-BY-SA 4.0 | available with the article |
| Automated Image Capture | Executable Files (.exe) | CC-BY-SA 4.0 | available with the article |
| Readme(Scanner) | Instructions for Scanner Use (.docx) | CC-BY-SA 4.0 | available with the article |
| BacteriaTrackerFiles | Matlab files (.m) | CC-BY-SA 4.0 | available with the article |
| Readme(BacteriaTrackingSteps) | Instructions for Bacteria Tracking (.docx) | CC-BY-SA 4.0 | available with the article |

**Bill of materials**

The Bill of Materials correspond to the acrylic sheets, tubing, and hinges to construct the chambers for the scanner and heater unit. There is an aluminum duct that provide the passage for steady air flow and a constant pre-set temperature through the chambers.

Bill of materials

| Designator | Component | # | Cost per unit –USD | Total cost -USD | Source of materials | Material type |
|---|---|---|---|---|---|---|
| Incubator (all-in-one digital thermostat, heater, fan control) | IncuKit™ XL for Reptile Incubators | 1 | $85.99 | $86 | https://incubatorwarehouse.com/incukit-xlrk.html | Electrical Units, Sensors |
| Acrylic sheets for incubation chamber | 12″ x 9.5″ x 3/8″<br>12″ x 12″ x 3/8″<br>11.25″ x 9.5″ x 3/8″<br>17.75″ x 9.5″ x 3/8″<br>25″ x 9.5″ x 3/8<br>25″ x 17.75″ x 3/8″<br>27″ x 10.5″ x 3/8″<br>20″ x 10.5″ x 3/8″<br>27.75″ x 20″ x 3/8″ | 2<br>1<br>2<br>2<br>2<br>1<br>2<br>2<br>1 | $90 | $90 | https://countryplasticsia.com/homepage/shop/ | Acrylic |
| Acrylic Hinges | Hinges | 4 | $3 | $12 | https://countryplasticsia.com/homepage/shop/ | Acrylic |
| Acrylic tubing | Tubing (diameter 4″ and length 2″) | 4 | $6 | $24 | https://countryplasticsia.com/homepage/shop/ | Acrylic |
| Aluminum Duct | 4-in. × 8 ft. flexible dryer vent duct | 1 | $10 | $10 | http://homedepot.com | Aluminum |
| Screws/Nails | 1/4 in.-20 × 1/2 in. Combo Round Head Zinc Plated Machine Screw | 48 | $0.30 | $14.40 | http://homedepot.com | Stainless Steel |
| EPSON Perfection V19 | Flatbed Scanner (4800 dpi) | 1 | $69.99 | $69.99 | http://adorama.com | Imaging Unit |

**Build instructions**

To build the bacterial growth and imaging hardware, the acrylic sheets are received from a machine shop and sorted according to their dimensions listed in Bill of Materials. First, the structure of the side enclosure is assembled as shown in Fig. 3A. To hold the structure in place, holes are drilled at the wall junctions of the side enclosure using an electric drill and screws are inserted in the holes to construct the side enclosure. Then, the heater/fan unit and acrylic tubing are manually put in the opening of the side enclosure as shown in Fig. 3B. Next, the structure of the interior scanner enclosure is assembled, holes are drilled in the interior scanner enclosure, and screws are inserted in the holes to construct the interior scanner enclosure. Thereafter, the interior scanner enclosure is attached to the side enclosure as shown in Fig. 3C. Next, the structure of the exterior scanner chamber is assembled, holes are drilled in the exterior scanner chamber, screws are inserted in the holes to construct the exterior scanner chamber. The exterior scanner chamber is then placed over to the interior scanner enclosure as shown in Fig. 3C. Two aluminum ducts are cut to suitable size and connected between the two openings of the side enclosure and those of the exterior scanner chamber for air flow as shown our Fig. 1A. The EPSON Perfection V750 Pro Scanner (or the EPSON Perfection V850 Pro Scanner) is placed within the interior scanner chamber and connected to a desktop computer. The power supply is connected to the heater/fan unit and scanner. Two cheaper versions of the EPSON scanner with a 4800 dpi resolution are the Epson Perfection V19 and V39 Flatbed scanner that cost $70 and $100, respectively.

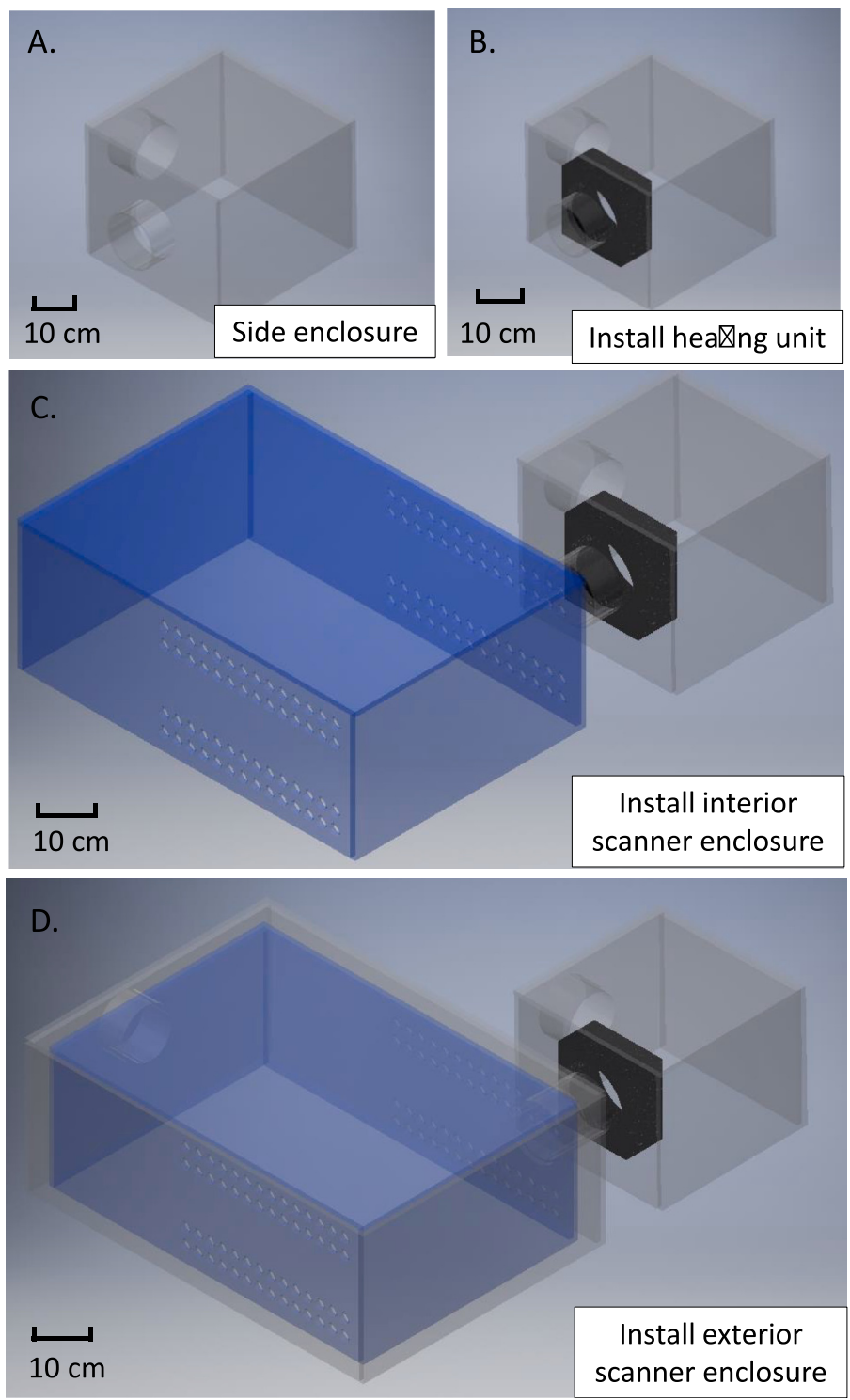

**Fig. 3.** CAD model assembly of the open-source hardware. (A) Hold the side enclosure in the upright position shown here. (B) Affix the heater/fan unit to the side enclosure. (C) Connect and install the interior scanner enclosure to the side enclosure as shown in the image. (D) Install the exterior scanner enclosure over the perforated interior enclosure.

**Operation instructions**

At the start of the experiment, the chambers and scanner are ensured to be clear of any unnecessary items, such as plates from previous experiments. The nutrient agar plates are prepared. The bacterial cultures (*Escherichia coli* wild-type), previously grown overnight, are diluted to approximately 50 to 100 CFU per mL. The controller of the heater/fan unit is turned on and the desired temperature is set for 37 °C. The temperature probe is placed inside the interior scanner enclosure. The chamber temperature reaches 37 °C after a wait time of around 35 to 40 min. The wild-type (mg1655) *E. coli* cells are plated as a cell suspension separately on six 3-inch nutrient agar plates using a pipette, and placed on the scanner. A custom .exe program is run, along with the Epson Scan Utility software, to automatically repeat the scanning operation every 5, 15 or 30 min for a total of 20 h, while saving the recorded image of six plates to a file folder. The .exe program allows users to select specific areas in the scanned A4-size image of six plates, and save them separately into a destination folder as six separate images of individual plates. This .exe custom program is written using the 'AutoIt' scripting language and available in a zip folder ("Automated Image Capture"). At the end of the experiment, the heater unit is turned off. Thereafter, the agarose plates are removed from the scanner, opened off their lids, sprayed with 70% ethanol to kill the bacteria, and disposed of in a bio-waste disposal unit. It is important to turn off the scanner and heater/fan unit immediately after the experiment to minimize equipment burn out. The heater/fan unit may break down after prolonged use, but affordable replacements are readily available from the manufacturer (mentioned in the Bill of Materials).

There are potential safety concerns with the hardware components. It is advisable not to physically touch or come close to the heater/fan unit or the vent system during or after its operation because its hot surface may damage the skin. The electrical wiring must be securely taped outside the chambers and away from the common space for handling. It is advised to open/close the acrylic chamber slowly and carefully to avoid physical injuries. All biosafety protocols should be followed while handling the nutrient agar plates with bacterial colonies. For working with nonpathogenic strain of E. coli, Biosafety Level 1 (BSL-1) protocols need to be followed. Key points to consider here are: (i) cleaning and disposing the agar plates after the experiments, (ii) turning off the heater and/or scanner after the experiments, and (iii) frequently checking that the image files are being saved properly and the scanner operations are working well.

**Validation and characterization**

After constructing the chamber housing, the heater/fan unit was set at 37 °C. The internal temperature within the chambers was observed monitored every few minutes for three separate days (Fig. 1 B). We found that the internal temperature was maintained at the pre-set temperature throughout the monitoring period. The heater fan ensured that the temperature was uniform throughout the two chambers. Trial runs were conducted to determine the optimum concentration of bacteria required for seeding on the plates. Time-scanned images of representative nutrient agar plates comprising the wild-type *E. coli* are shown in Fig. 4. The saved images of agarose plates were run through our custom MatLab programs (listed in

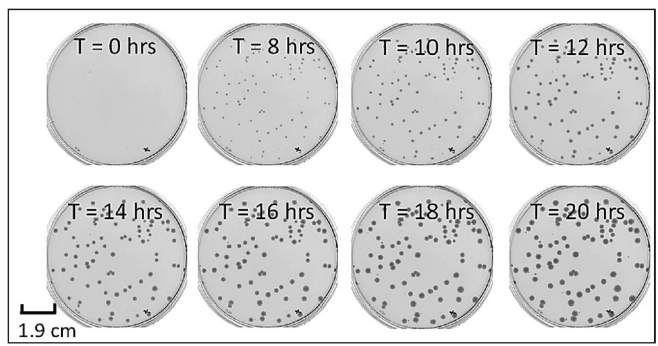

**Fig. 4.** Growth and imaging of wild-type E. coli with our system. The colonies are noticeable within less than 8 h and the colony area grows with time for five different sets of experimental runs (N = 5).

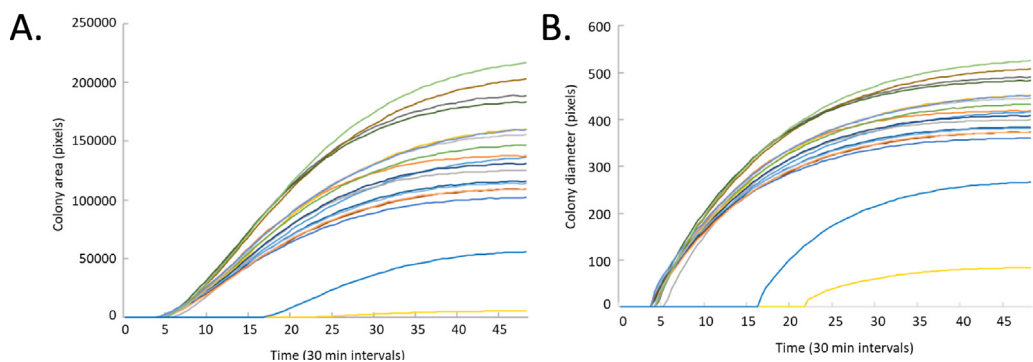

**Fig. 5.** Lag time and growth kinetics of bacterial colonies. (A, B) The colonies of wild-type *E. coli* emerge within 5 to 8 h of plating with an exponential rise and eventual saturation phase.

Design Files Summary) to extract the area of different bacterial colonies at different time points. Almost all the wild-type colonies emerge within 5 to 8 h of plating on agarose plates (Fig. 5).

**Limitations and future scope**

Our open-source station can be constructed with minimal costs to conduct experiments in microbiology where automated tracking the growth of bacterial colonies is made possible. Purchasing commercial setups for this purpose may not be cost-effective and are often closed source for any system modifications. However, there are certain limitations of the presented technology. As with most experiments in bacteriology, the natural heterogeneity in different cell populations makes it challenging to standardize the image processing steps for all experiments [3]. At the start of experiment, it is difficult to estimate the optimal dilution to start with, but we have seen that a low to mid CFU per mL (i.e. 50 to 100) helps to prevent grouping of cells or merging of colonies. Only a finite number of bacterial colonies (less than 250 CFU per mL [15]) can be tracked over time within the maximum scanning area, and the imaging has to be completed before the colonies merge or start competing for nutrients. It is difficult to ascertain the actual number of cells in each colony by scanning plate images. Single cell studies are not possible with this method but can be performed using microfluidic chips at a lower dilution (Fig. 6). Finally, considering that each image file may be over 10 Mb, there is a finite number of high-resolution images that can realistically be captured by the scanner during each experiment.

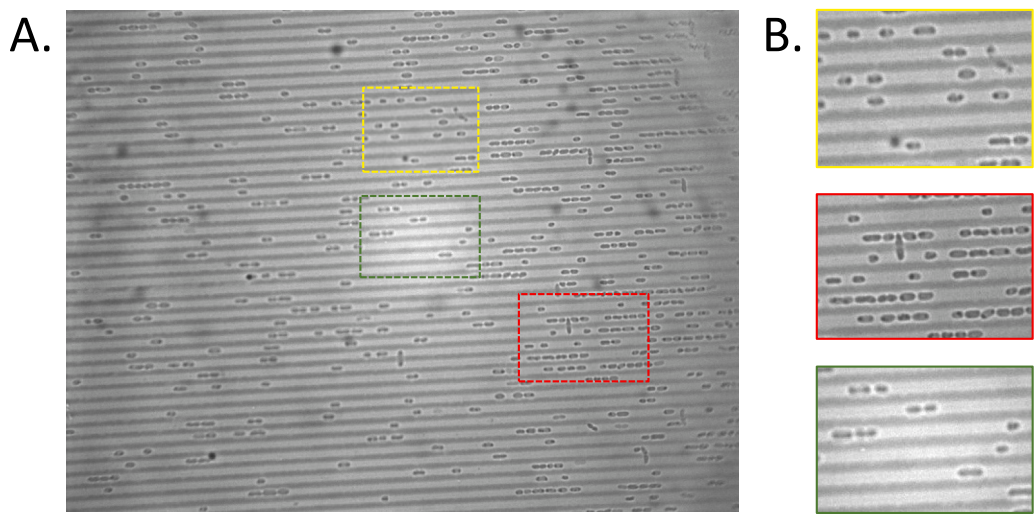

**Fig. 6.** Microfluidic chips to quantify elongation and division of single bacterial cells. (A) Microfluidic channels were fabricated to hold multiple discrete rows of single cells with slow perfusion of nutrient media. (B) Snapshots of different sections of the microfluidic chip, showing the elongation and cell division with adequate resolution. These single cell studies complement the population-scale quantification of colony growth and lag time.

In the near future, there is scope to expand the imaging functionalities to unravel new phenotypic traits and characterize the natural heterogeneity in bacterial populations. For example, our method can be adapted to probe other characteristics of bacterial colonies, such as their morphology, organization, and/or color for addressing unanswered questions in biofilm formation and antibiotic resistance. Persistence is still poorly understood and there are many attributes that could be quantified by the measuring the lag time and growth kinetics of different persistors under different stressful conditions. The results obtained from our scanner-based imaging method can further be validated with conventional tools in microbiology, such as flow cytometry, optical plate readers, turbidity measurements, and single cell analysis. An example of open source hardware and software for the automated counting of bacterial colonies showed that high-resolution images from flatbed scanners (1200 dpi and higher) were superior than those from digital cameras for bacterial enumeration [16]. Besides applications in microbiology, our bacterial enumeration system can be diversified for imaging, counting, and tracking other microorganisms on agar plates, such as *C. elegans* and parasitic nematodes [17–20]. For this purpose, a number of open source image processing and analysis techniques are freely available through the ImageJ software and the scientific imaging community [14,21,22]. Finally, the presented system offers a range of possibilities for STEM education and direct hands-on training for undergraduate and graduate students who want to tinker with the system and methods to optimize its performance and throughout at a low cost–something that is difficult with commercial cell trackers. For example, students can learn to prepare microbiology experiments, construct incubation chambers from CAD drawings and off-the-shelf components, and write open source software to control the scanner operations and track colonies over time.

## CRediT authorship contribution statement

**Santosh Pandey:** Conceptualization, Methodology, Validation, Investigation, Writing – review & editing, Funding acquisition, Software, Writing – original draft, Visualization, Data curation, Supervision, Resources. **Yunsoo Park:** Conceptualization, Methodology, Validation, Investigation, Writing – review & editing, Funding acquisition, Software, Writing – original draft, Visualization, Data curation, Supervision, Resources. **Ankita Ankita:** Conceptualization, Methodology, Validation, Investigation, Writing – review & editing, Funding acquisition, Software, Writing – original draft, Visualization, Data curation, Supervision, Resources. **Gregory J. Phillips:** Conceptualization, Methodology, Writing – review & editing, Funding acquisition, Writing – original draft, Investigation, Visualization, Resources.

## Declaration of Competing Interest

The authors declare that they have no known competing financial interests or personal relationships that could have appeared to influence the work reported in this paper.

## Acknowledgements

This work was partially supported by the U.S. Defense Threat Reduction Agency (HDTRA1-15-1-0053) and U.S. National Science Foundation (NSF IDBR-1556370). We are grateful to all our past graduate students who worked on the bacterial imaging system, including Zach Njus, Upender Kalwa, Taejoon Kong, Nicholas Backes, Augustine Beeman, Jared Jensen, Elizabeth Wlezien, and Christopher Legner. We also thank Dr. Gregory Tylka and his laboratory personnel for improving the system functions to image populations of soybean cyst nematodes.

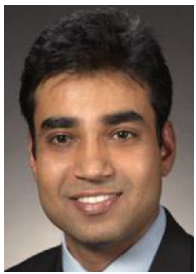

**Santosh Pandey** is an Associate Professor in the Department of Electrical and Computer Engineering at Iowa State University. He supervises the Micro/Nano Systems Laboratory in Coover Hall where research projects are conducted in experimental areas of sensors, microfluidics, instrumentation, imaging, and software processing, and data analytics.

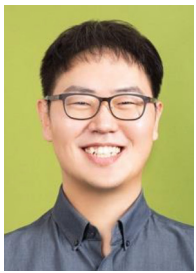

**Yunsoo Park** is a Master Student at Iowa State University. His research interests include machine learning, imaging, sensors, low-powered PCB design and low-powered wireless communication. He participated in the Micro/Nano Systems Laboratory in Coover Hall.

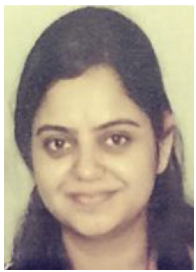

**Ankita** is a Ph.D. student at Iowa State University. She works in the Micro/Nano Systems Laboratory in Coover Hall. Her research interests include bioengineering, machine learning, low-power PCB design, video analytics and image processing.

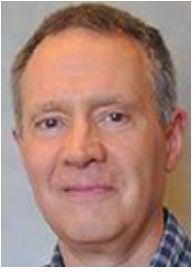

**Gregory Phillips** is a Professor in Veterinary Microbiology and Preventive Medicine at Iowa State University. His research interests are in genomics and metagenomics, microbiome/host interactions, antibiotic persistence in bacterial pathogens, and bacterial membrane protein localization. He received the Pfizer award for research excellence in veterinary medicine. He is in the editorial board of Plasmid, Journal of Bacteriology, and EcoSal.